\documentclass[twoside,12pt,onecolumn]{article}
\usepackage{hyperref}
\usepackage{authblk}
\usepackage{cite}
\usepackage{graphicx}
\usepackage[margin=0.8in]{geometry}
\usepackage{helvet}
\usepackage{wrapfig}
\usepackage{float}

\title{Racial/ethnic and socioeconomic disparities of Covid-19 attacks rates in Suffolk County communities.}
\author[1]{Daniel Dobin}
\author[2]{Alexander Dobin  \thanks{dobin@cshl.edu}}
\affil[1]{Candlewood Middle School, New York, USA}
\affil[2]{Cold Spring Harbor Laboratory, New York, USA}
\date{}

\begin{document}
\maketitle

\begin{abstract}
We investigated the dependence of Covid-19 attack rates on demographic and socioeconomic factors for the communities in Suffolk County (Long Island, New York State), presently the 5th most-affected county in the United States. Confirming the previous observations that minorities are disproportionately impacted by the Covid-19 disease, we found that the attack rate is strongly correlated with the minority population proportion, with an alarmingly high $\sim4$-fold attack rate increase for Black and Hispanic populations.
\end{abstract}

\section{Introduction.}
The Covid-19 disease first emerged in late 2019 and has since risen to a global health crisis and was declared a pandemic on March 11th, 2020. Covid-19 is caused by the Severe Acute Respiratory Syndrome Coronavirus 2 (SARS-CoV-2) which is transmitted from person to person in close contact by small respiratory droplets. As of \today, this highly contagious disease has infected more than 2 million people and has caused over 150 thousand deaths in more than 200 countries.

Socioeconomic and racial/ethnic inequities frequently affect health and healthcare access, resulting in a higher burden of disease and mortality in vulnerable social groups \cite{Braveman2010,Flores2010,Nesbitt2016,Williams2016}.
Previous studies showed that multiple diseases, both non-infectious (e.g. cardiovascular diseases, cancer, obesity, diabetes) and infectious (influenza, tuberculosis, pneumonia), have a higher incidence rates in the minority populations \cite{Graham_heartDispar_2015,Ozdemir_cancerDispar_2017,Bhupathiraju_obesityDiabetes_2016,Quinn_influenza_2011,_tuberculosis_2004,Burton_pneumonia_2010}.

There have been multiple reports of significant impacts of racial/ethnic disparities on the severity of Covid-19. For instance, the CDC COVID-NET survey of 580 hospitalized Covid-19 patients from 14 US states \cite{COVID-NET_2020} found that $33\%$ of the patients were African Americans, which is significantly higher than the  percentage of African Americans ($18\%$) in the COVID-NET catchment population. In New York City, African Americans account for $28\%$ of all fatalities while accounting for 22\% of the total population, and Hispanic Americans account for $34\%$ of the total cases while accounting for 29\% of the population \cite{NYS_COVID19_tracker} .

Measuring the effects of socioeconomic and racial disparities can help inform the public and policymakers about at-risk populations, resulting in a more efficient allocation of resources and personnel to effectively mitigate the spread of Covid-19. It could also benefit the most affected communities by increasing their awareness of the Covid-19 dangers and preventative measures.

In this work, we investigated the dependence of Covid-19 attack rate on racial/ethnic composition and several socioeconomic metrics for the communities in Suffolk County (Long Island, New York state), presently the 5th most affected county in the United States. We found that the attack rate is strongly correlated with the minority population percentage, confirming the previous observation that minorities are much stronger affected by the Covid-19 disease.

\section{Methods.}
We have collected the Covid-19 attack rate data on the Suffolk county Department of Health Services interactive \href{https://gis.suffolkcountyny.gov/portal/apps/opsdashboard/index.html#/76a26a0c83634266aa9efc35bd4f1975}{web-map} on April 10-11 \cite{SuffolkWebMap}.
The attack rate is defined as the total number of positive cases in a community (at the collection time) divided by the total population of the community. Next, we collected demographic and economic data for each community from the \href{censusreporter.org} {censusreporter.org} website \cite{CensusReporter}.
This information was sourced from the \href{https://www.census.gov/newsroom/press-releases/2019/acs-5-year.html}{2014-2018 American Community Survey} \cite{ACS}.  The following metrics were considered: proportions of White, Black, Hispanic and Asian residents, median age, per capita income, median household income, proportion of persons below poverty level, mean travel time to work.
We restricted our analyses to the 128 communities with populations greater than 1,000 persons. We utilized standard statistical and plotting functions in Matlab; confidence intervals for all estimates were calculated by bootstrapping with 10,000 samples. The data table and code are available on GitHub: \url{https://github.com/alexdobin/Covid19_disparities}.

\section{Results.}

\begin{figure}[h]
  \begin{center}
  \caption{Covid-19 attack rate vs. minority population percentage. The circles represent communities, the solid red line is a linear least squares fit to the data, the dashed magenta lines represent the 95\% confidence bands of the fit obtained by bootstrapping (10,000 samples).}

  \label{fig:AttackRateVsMinority}  	
  
  \textbf{Figure 1}

  \includegraphics[width=0.6\linewidth]{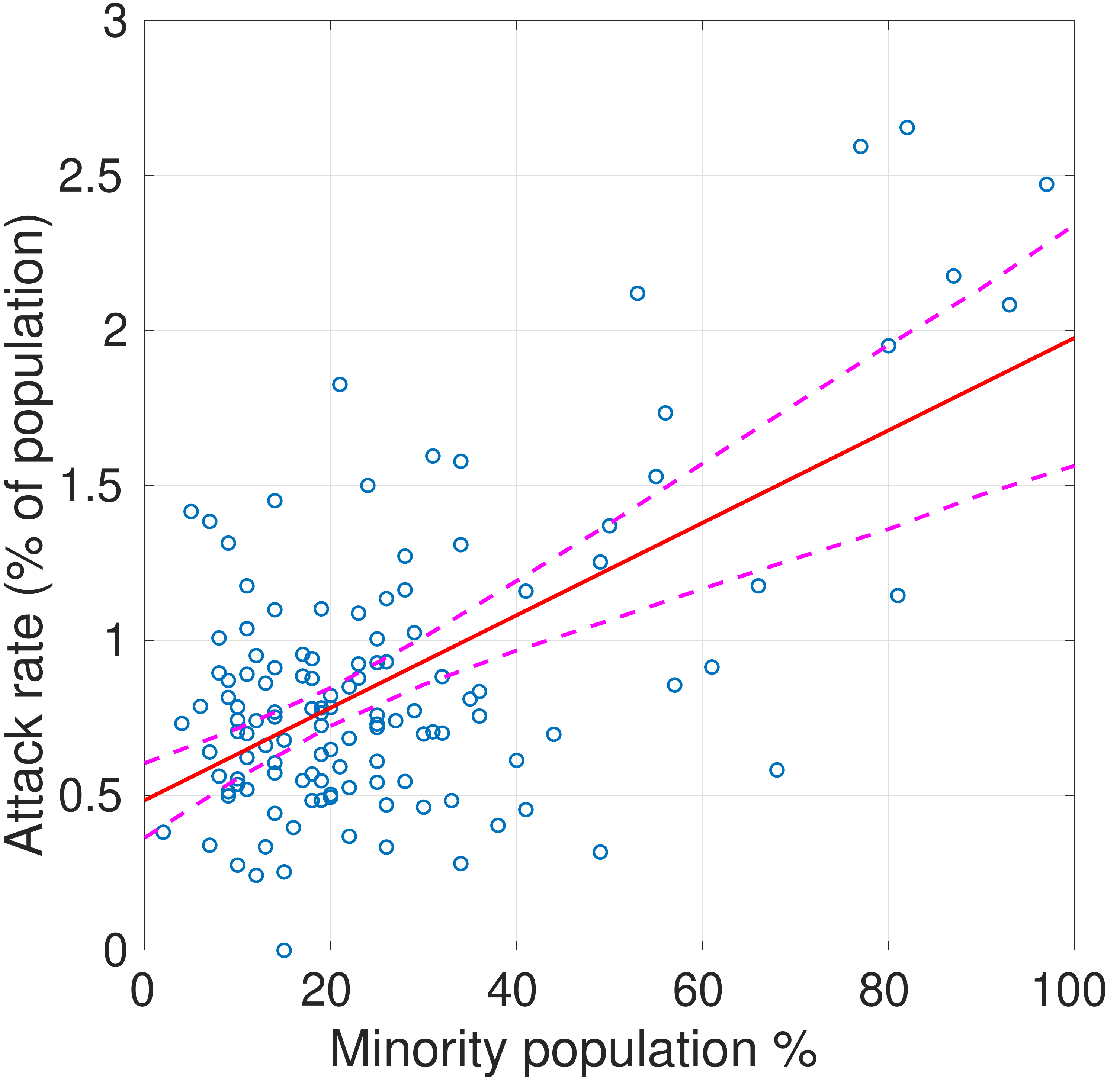}
  \end{center}

\end{figure}

In Figure \ref{fig:AttackRateVsMinority} we plotted the Covid-19 attack rate vs. the minority population percentage in each community.
Attack rate correlates strongly with the minority population proportion: Pearson correlation coefficient $R=0.61$,
$95\%$ confidence interval $[0.41,0.75]$.
Performing the linear least-squares fit to the data, we estimate that the attack rate for the minorities is 4.1 times higher than that for the whites ($95\%$ confidence interval $[2.6,6.1]$).
We note that this ratio is higher than ratios that can be derived from the previous reports of rascal/ethnic composition of fatalities in New York City
$(28\%+34\%)/(22\%+29\%) \approx 1.2$
\cite{NYS_COVID19_tracker} and hospitalized patients across the US $ 33\%/18\% \approx 1.8 $ \cite{COVID-NET_2020}.

\begin{figure*}
   \caption[Attack rate vs Income]
   {A: attack rate vs. per capita income. Circles represent communities, their color scale corresponds to the percentage of the minority population in each community.

   B: Per capita income vs. minority population percentage. The curved red line represents the quadratic least squares fit.
   }
   \label{fig:AttackRateVsIncomePerCapita}

  \textbf{Figure 2A}

  \vspace{-8mm}
  \begin{center}
     \includegraphics[width=0.64\linewidth]{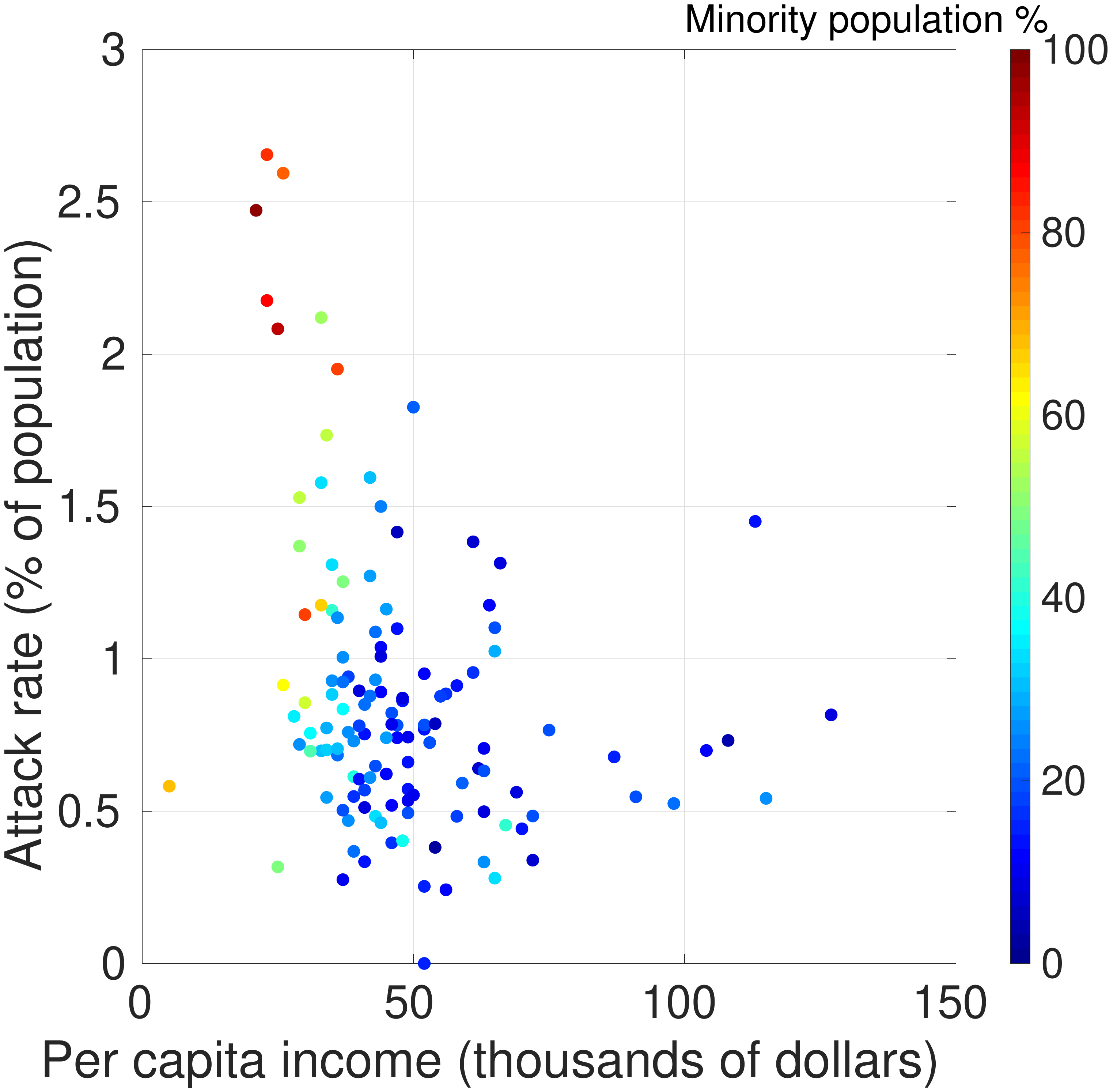}
  \end{center}
  \textbf{Figure 2B}

  \vspace{-14mm}
  \begin{center}
     \includegraphics[width=0.64\linewidth]{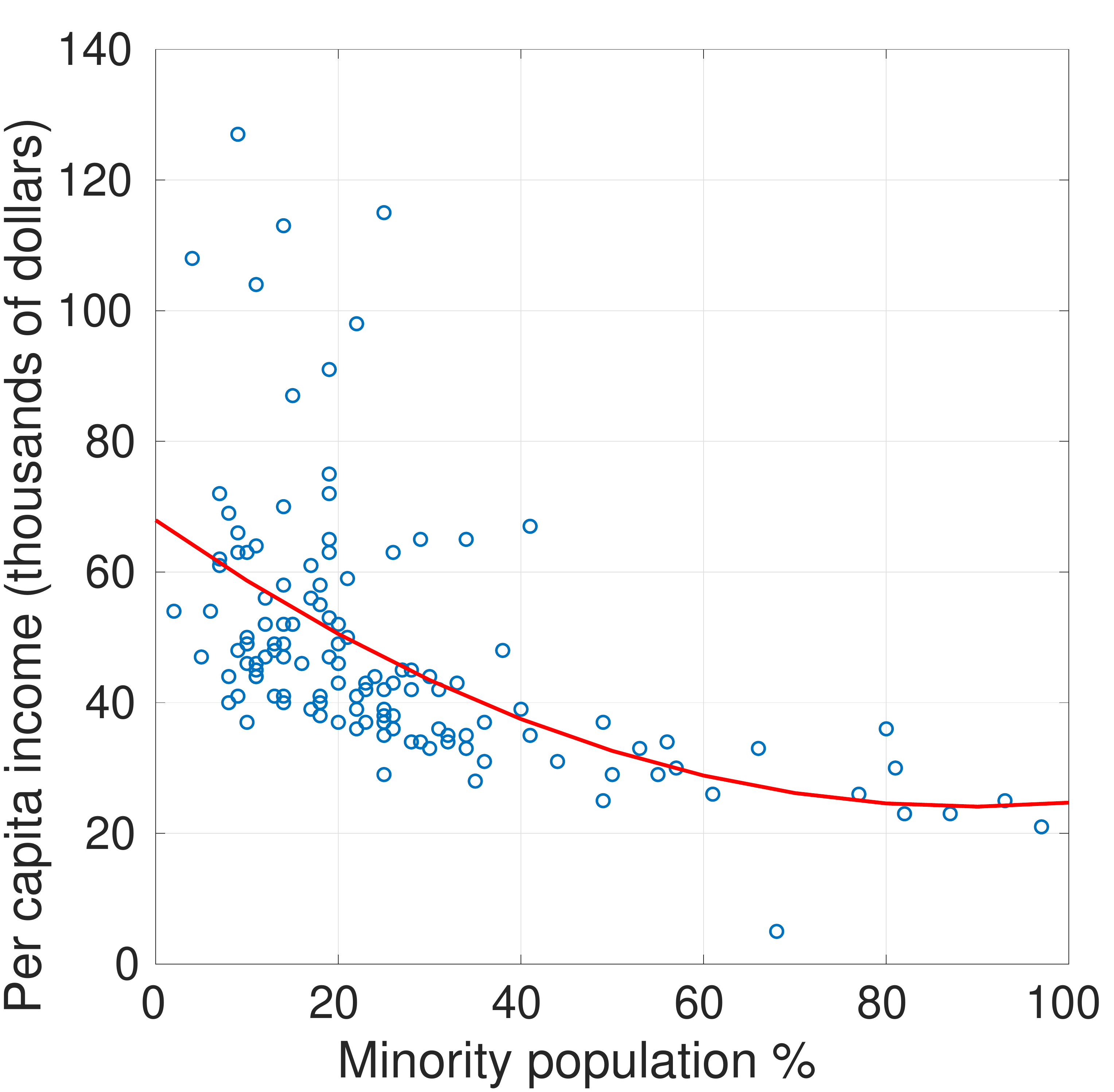}
   \end{center}
\end{figure*}

In Figure \ref{fig:AttackRateVsIncomePerCapita}A we plotted Covid-19 attack rate vs. per capita income.
We observe that low-income communities have higher attack rates than high-income communities.
However, as shown in figure 2B, the per capita income is itself strongly anti-correlated with the minority population proportion.
For communities with minority population $<20\%$, we found no significant correlation between attack rate and per capita income: $R=0.14$, $95\%$ confidence interval $[-0.06,0.45]$.

\begin{figure*}
    \caption[Correlations with multiple predictors]
    {A: Matrix of pairwise Pearson correlation coefficients between the predictor variables and attack rate.

    B: Pearson correlation of attack rate with different predictors. Error bars represent the 95\% confidence intervals calculated using bootstrapping (10,000 samples).
    }
    \label{fig:CorrelationMatrix}
    \textbf{Figure 3A}

    \vspace{-8mm}
    \begin{center}
        \includegraphics[width=0.75\linewidth]{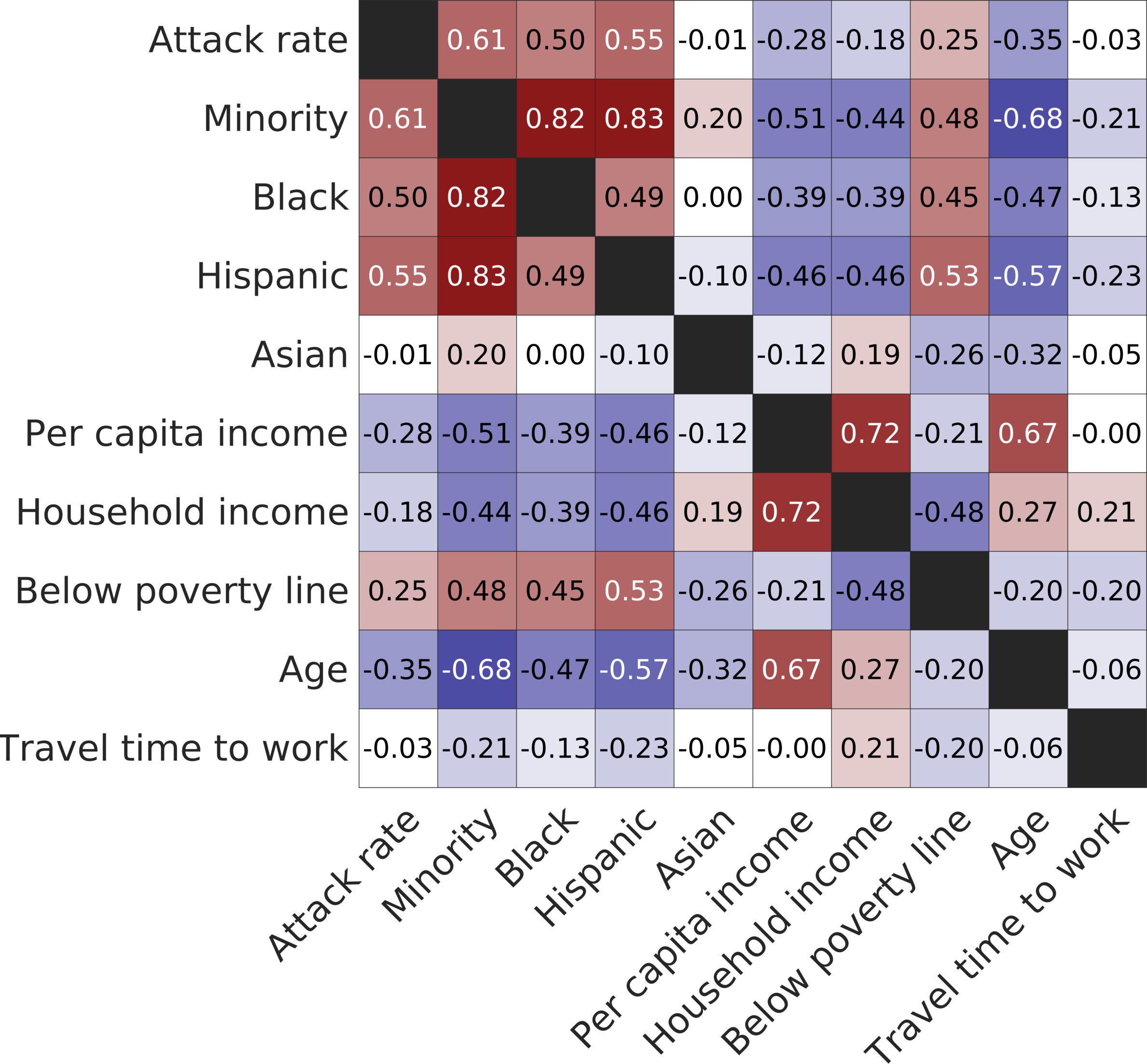}
    \end{center}
  \textbf{Figure 3B}
  \vspace{-1cm}
  \begin{center}
      \includegraphics[width=0.5\linewidth]{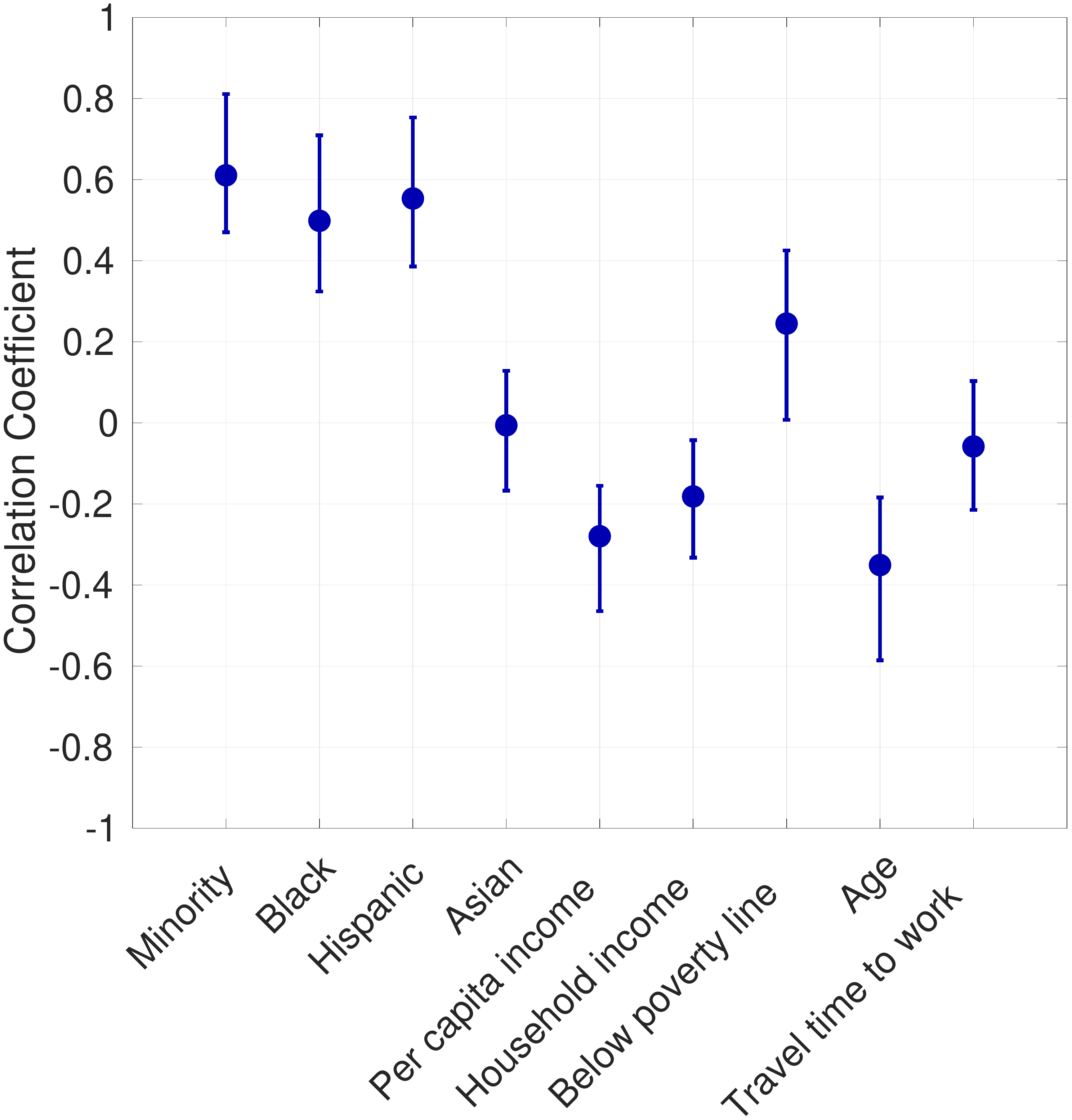}
  \end{center}

\end{figure*}

The multicollinearity is also present for other demographic and economic predictor variables in our study.
Figure \ref{fig:CorrelationMatrix}A shows pairwise correlation coefficients between the predictor variables, as well as attack rate and predictor variables.
We see a strong negative correlation of minority population percentage with per capita income, mean household income, and age, and a positive correlation with proportion of persons below the poverty line, and the same trends are observed in the dependence of attack rate on these predictors.
Figure \ref{fig:CorrelationMatrix}B shows that the minority population proportion has the highest absolute correlation with the attack rate. The attack rates are correlated with Black and Hispanic population proportions equally strongly. The Asian population proportion and travel time to work do not correlate significantly with the attack rate, though this may be owing to the narrow range of these variables in the communities under study.
We also investigated multilinear regression of the attack rates with all pairs of predictor variables, and found that none of them increased the correlation statistically significantly, which indicates that our study is not powered to discern separate contributions of these demographic and socioeconomic variables.

\section{Discussion.}
In this study, we investigated the dependence of the Covid-19 attack rate on the several demographic and economic factors in the Suffolk County communities. Among all the variables, the percentage of the minority population in a community was the best predictor of Covid-19 attack rate. Unlike other approaches, which utilize the demographics of the infected patients, our study estimates the attack rate dependence on race/ethnicity by correlating it with the racial/ethnic composition of communities. While the former approaches are more direct, our approach is more informative for assessing hardships and disparities affecting the minority communities.

Our work does \textbf{not} utilize any genetic information and hence does \textbf{not} point to any genetic basis in the racial/ethnic disparities of the Covid-19 infection. The racial/ethnic and socioeconomic inequities have been widely documented for other diseases and are attributed to a multitude of health and environmental causes, such as limited access to healthcare, lack of health insurance, pre-existing conditions, substandard housing, and hazardous work. In the case of Covid-19 pandemic, these hardships are further exacerbated by the higher economic stress for minority communities resulting in diminished capacity for social distancing which is imperative to suppress the spread of the novel coronavirus. In addition, minorities constitute a large proportion of essential workers, further increasing their exposure to Covid-19 infection.

Our main finding is that the Covid-19 attack rate for Black and Hispanic populations is alarmingly (4.1-fold) higher than that for the White population. In line with previous observations, this indicates that the minorities are affected the greatest by Covid-19, and thus minority communities require the most assistance in order to lessen the distress caused by the Covid-19 pandemic. We notice that the New York State and Suffolk County governments have begun taking first steps to address this crisis. The new Covid-19 testing sites have been opened on April 8-16  in minority communities that are Covid-19 hot spots: Brentwood (3-fold higher attack rate than Suffolk County average), Huntington Station (2.4-fold), Wyandanch (2.8-fold). We hope that our study will increase public and government awareness of the severity of the Covid-19 impact on minorities, and result in more decisive actions taken to protect the minority communities in the United States and around the world.

\newpage
\bibliographystyle{unsrt}
\bibliography{Covid19_disparities}

\newpage
\section*{Figures}
\clearpage

\clearpage

\end{document}